\documentstyle[12pt]{article}
\newcommand{\secn}[1]{Section~\ref{#1}}

\newcommand{\eq}[1]{Eq.~(\ref{#1})}

\newcommand{\nl}{\nonumber \\}

\def\beq{\begin{equation}}
\def\eeq{\end{equation}}
\def\beqa{\begin{eqnarray}}
\def\eeqa{\end{eqnarray}}
\newcommand{\sect}[1]{\setcounter{equation}{0}\section{#1}}
\renewcommand{\theequation}{\thesection.\arabic{equation}}
\newcommand{\EQ}{\begin{equation}}
\newcommand{\EN}{\end{equation}}
\newcommand{\bea}{\begin{eqnarray}}
\newcommand{\ena}{\end{eqnarray}}

\renewcommand{\a}{\alpha}
\renewcommand{\b}{\beta}

\newcommand{\e}{\epsilon}

\newcommand{\NP}[1]{Nucl.\ Phys.\ {\bf #1}}
\newcommand{\PL}[1]{Phys.\ Lett.\ {\bf #1}}

\newcommand{\PR}[1]{Phys.\ Rev.\ {\bf #1}}

\renewcommand{\thefootnote}{\fnsymbol{footnote}}
\def\one{{\hbox{ 1\kern-.8mm l}}}

\begin{document}
\begin{titlepage}
\rightline{DFTT 40/97}
\rightline{NORDITA 97/49-P}
\rightline{\hfill July 1997}
\vskip 1.2cm
\centerline{\Large \bf CLASSICAL $p$-BRANES FROM BOUNDARY STATE
\footnote{Work partially supported by the European Commission
TMR programme ERBFMRX-CT96-0045 in which R.R. is associated to
Torino University.}}
\vskip 1.2cm
\centerline{\bf Paolo Di Vecchia\footnote{e-mail:
DIVECCHIA@nbivms.nbi.dk}}
\centerline{\sl NORDITA}
\centerline{\sl Blegdamsvej 17, DK-2100 Copenhagen \O, Denmark}
\vskip .2cm
\centerline{\bf Marialuisa Frau, Igor Pesando and Stefano
Sciuto}
\centerline{\sl Dipartimento di Fisica Teorica, Universit\`a di
Torino}
\centerline{\sl Via P.Giuria 1, I-10125 Torino, Italy}
\centerline{\sl and I.N.F.N., Sezione di Torino}
\vskip .2cm
\centerline{\bf Alberto Lerda\footnote{II Facolt\`a di Scienze
M.F.N., Universit\`a di Torino
(sede di Alessandria), Italy}}
\centerline{\sl Dipartimento di Scienze e Tecnologie Avanzate and}
\centerline{\sl Dipartimento di Fisica Teorica, Universit\`a di
Torino}
\centerline{\sl Via P.Giuria 1, I-10125 Torino, Italy}
\centerline{\sl and I.N.F.N., Sezione di Torino}
\vskip .2cm
\centerline{\bf Rodolfo Russo}
\centerline{\sl Dipartimento di Fisica, Politecnico di Torino}
\centerline{\sl Corso Duca degli Abruzzi 24, I-10129 Torino, Italy}
\centerline{\sl and I.N.F.N., Sezione di Torino}
\vskip 1cm
\begin{abstract}
We show that the boundary state description of a D$p$-brane
is strictly related to the corresponding classical solution of
the low-energy
string effective action. By projecting the boundary state
on the massless states of the closed string we obtain the tension, the
R-R charge and the large distance behavior of the classical solution.
We discuss both the case of a single D-brane and that of bound
states of two D-branes.
We also show that in the R-R sector the boundary state, written
in a picture which treats asymmetrically the left and right
components, directly yields the R-R gauge potentials.
\end{abstract}
\end{titlepage}
\newpage
\renewcommand{\thefootnote}{\arabic{footnote}}
\setcounter{footnote}{0}
\setcounter{page}{1}
\sect{Introduction}
\label{intro}
Classical solutions of various low-energy string actions carrying a
non-vanishing charge with respect to some $(p+1)$-form field have been
intensively studied in the last few years. The simplest of them are
discussed in detail in Ref.~\cite{DUFF} where one can also find
the references to the original papers.
The classical solutions having a
non-vanishing electric or magnetic charge under the
Neveu-Schwarz--Neveu-Schwarz (NS-NS) $2$-form correspond,
respectively, to the fundamental string and the solitonic $5$-brane, or,
in the dual formulation, to the solitonic string and
the fundamental $5$-brane. On the contrary,
the classical solutions with
a non-vanishing charge under the various
$(p+1)$-forms of the Ramond-Ramond (R-R)
sector do not have any relation to the
perturbative closed string or its solitons. In fact,
as it has been recognized by Polchinski~\cite{POLC}, these solutions which
are required by various string dualities, correspond to membranes
on which open strings can end with
Dirichlet boundary conditions in the transverse directions and
the usual Neumann boundary conditions in the longitudinal
directions. For this reason they are called Dirichlet branes (D-branes).
The properties of D-branes~\footnote{For a
review of their properties see Ref.~\cite{LPOLC}.} can be
discussed either by studying their interaction with open
strings~\cite{POLC,BACHAS,Kleb} or, more
efficiently, by introducing the so-called boundary
state~\cite{NBOUND,BILLO,FRAU,OTHERS}~\footnote{For a pre-D-brane discussion
of the boundary state see Refs.~\cite{CREMMER,bounstate}.}.

The boundary state is a BRST invariant state
written in terms of the closed string oscillators
which basically identifies the left and the right sector of the closed
strings attached to it. It contains the couplings of the
D-brane with all the states of the closed string spectrum and, in
particular, as we
show in Sections 2 and 4, it can be used to compute the
coupling with massless states, providing an independent
calculation of the D-brane tension and R-R charge~\cite{POLC}.

In this paper we show that the boundary state can also
be used to efficiently
extract target-space information, such as the value of background
fields corresponding to the classical description of a
D-brane. We show that
the boundary state formalism greatly simplifies the calculation
of the long range fields produced by a flat membrane presented in
Refs.~\cite{Kleb,Myers} where the string scattering from a D-brane
was studied in detail and then, in the low energy limit, the various
amplitudes were factorized  in a pure tree graph and in a source term
generated by the presence of the D-brane.
However, this factorization can be
performed once for all at string
level and, indeed, it can be seen as the very definition of the boundary
state~\cite{FRAU}. Then the long-distance behavior of the various fields
present in the low energy string action can be deduced by studiyng
the emission of massless string states from the boundary state.
This shows that there is a very direct
connection between the classical D-brane solution and the corresponding
boundary state describing its quantum properties.
While for the NS-NS sector the calculations in the case of
superstring
are very similar to the ones performed for the bosonic string, in the
R-R sector we see that the expression for the boundary state has a
particularly simple expression if it
is written in an unusual picture which treats asymmetrically the left
and the right sector.
The same analysis can be performed also in the case of
D-brane bound states, again finding complete agreement with the
classical solutions.

The paper is organized as follows. In \secn{sez2}, in order to
avoid all the technicalities of the superstring, which at first
sight may obscure the otherwise very clear physical picture, we
concentrate on the bosonic string.
In this theory we write the boundary state
for a D$p$-brane and show how to extract from it
the brane tension that appears in the effective description
provided by the Born-Infeld action.
Then, after a short discussion
of the classical D-brane solution, we derive its large-distance
behavior and
compare it with what we get from the boundary state by
studying the emission of
the graviton, dilaton and antisymmetric Kalb-Ramond field. As expected,
we find complete agreement between the two approaches
provided the space-time
dimension is ten. This forces us to consider the superstring case.
In \secn{sez3} we write the boundary state for the NS-NS sector of the
superstring theory and show a complete
agreement between the large-distance
behavior of the classical solution and
that obtained from the boundary state. In \secn{sez4}
we write the boundary state
for the R-R sector of superstring and show that, when it is saturated
with states in the corresponding asymmetric picture, there is complete
agreement between the long-distance behavior of the classical solution
and the boundary state results, also in the case of the $(p+1)$-form
potential. Finally, in \secn{sez5} we turn to D-brane bound states
with two non-vanishing $p$-form fields whose dimensions differ by two.
They can be described by a boundary state containing a
non trivial background gauge field. Again we find a complete agreement
between the large-distance behavior obtained from the boundary state
and that following from the low energy solutions
described in Ref.~\cite{Myers2}.
Some technical details as well as our conventions and notations
for spinors in the R-R sector are contained in the Appendix.

\vskip 1.5cm
\sect{D-brane tension and large-distance behavior}
\label{sez2}

A D$p$-brane can be conveniently described in terms of a
string world-sheet with a boundary on which $d-p-1$ coordinates
satisfy Dirichlet boundary
conditions, $d$ being the space-time dimension.
In this framework a very useful object is the boundary
state $|B\rangle$
which is a BRST invariant state of the closed string
that inserts a boundary on
the world-sheet and
enforces on it the appropriate boundary conditions.
It was originally introduced in order to factorize open string
loop diagrams in terms of the closed string states~\cite{CREMMER},
and its explicit form can be obtained either by solving overlap
equations between the left and right fields
induced by the presence of a boundary~\cite{bounstate}, or by factorizing
amplitudes of closed strings emitted from a disk~\cite{Ad,FRAU}.

In order to see in the most transparent way how the boundary state
describes a D$p$-brane, we begin with the bosonic
string and, for the sake of
simplicity, understand the ghost degrees of freedom.
The boundary state for a D-brane located at $y$ is
\beq
|B\rangle ={\cal N} \delta^{(d_{\bot})} (q - y) ~
\exp \left[ - \sum_{n=1}^{\infty} a_n^{\mu\dagger}
{\cal S}_{\mu \nu} {\widetilde a}_n^{\nu\dagger}
\right]  | 0; k= 0\rangle~~,
\label{bounda0}
\eeq
where the $\delta$-function is only over the $d_\bot=d-p-1$
directions transverse to the brane,
and $a_n$ ($\widetilde{a}_n$)
are the left (right) moving modes of the string coordinates. The
matrix\footnote{Throughout this paper,
the indices $\alpha$ and $\beta$ denote
the $p+1$ longitudinal directions of the D$p$-brane, while the indices
$i,j$ denote the transverse directions.}
\beq
{\cal S}_{\mu \nu} = (\eta_{\alpha \beta}, - \delta_{ij} )
=2\,(\eta_{\alpha \beta},0) - \eta_{\mu \nu}
\label{S}
\eeq
is obtained from the metric $\eta_{\mu\nu}$ by performing a
T-duality transformation on the last $d_\bot$ directions
to enforce the Dirichlet boundary conditions
appropriate for a D$p$-brane.
The normalization constant ${\cal N}$ was
{\it derived}  in Ref.~\cite{FRAU} from the factorization of
amplitudes of closed strings emitted from a disk in
the case where each target-space direction was compactified
on a circle of radius $R$.
In the decompactification limit $R\to \infty$, the boundary state
normalization can be written as
\beq
{\cal N} = \frac{4}{\a'}  \sqrt{W\,\a' \over 8\pi V} \,
\left(2\pi\sqrt{2\a'} \right)^{-d/2} (2\pi R)^{d_\bot}~~,
\label{norm0}
\eeq
where $W$ is the normalization of the
open string momentum eigenstates
\beq
W = (2\pi R)^{p+1} \left(\frac{2\pi\a'}{R}\right)^{d_\bot}~~,
\label{W}
\eeq
while $V$ is the space-time volume $V=(2\pi R)^d$.
Note however, that the
factorization procedure naturally leads
to a boundary state containing the
closed string propagator
\beq
D_a = \frac{\alpha '}{ 4\pi}\, \int_{|z| \leq 1} \frac{d^2 z}{ |z|^2}
~z^{L_0 -a} {\bar{z}}^{{\widetilde{L}}_0 -a}~~.
\label{pro}
\eeq
with intercept $a=1$. Thus the expression
derived in Ref.~\cite{FRAU} is slightly different from
\eq{bounda0}, since the latter does not contain the closed string
propagator (\ref{pro}). For future use, it is convenient to rewrite the
normalization factor ${\cal N}$ in the following form
\beq
{\cal{N}} = \sqrt{\frac{2}{\alpha ' \pi}}
\left(2\pi\sqrt{2\a'} \right)^{-d/2}
\left(2 \pi \sqrt{\alpha '} \right)^{d_{\bot}}~~.
\label{norm1}
\eeq
The first two factors correspond to
the normalization in the pure Neumann case,
while the last one provides a factor of $( 2 \pi \sqrt{\alpha '})$ for
each transverse component of the
$\delta$-function present in the
boundary state in Eq. (\ref{bounda0}). This
observation will be used later on when
we construct the boundary state for a
non localized D-brane.

Using \eq{norm1}, we can rewrite the boundary state $|B\rangle$ as
\beq
|B\rangle =  \frac{{\widehat{T}}_{p}}{2} ~
\delta^{(d_{\bot})} (q - y) ~
\exp \left[ - \sum_{n=1}^{\infty} a_n^{\mu\dagger}
{\cal S}_{\mu \nu} {\widetilde a}_n^{\nu\dagger}
\right]  | 0; k= 0\rangle
\label{bounda}~~,
\eeq
where
\beq
{\widehat{T}}_{p} =
\frac{\sqrt{\pi}}{2^{(d- 10)/4}} \left(4 \pi^2 \alpha '
\right)^{(d - 2p -4 )/4}~~.
\label{nor}
\eeq
If the matrix element $ \langle B | D_{a=1} |B \rangle$ is computed,
one obtains the same result
derived by Polchinski~\cite{POLC} using the open string formalism
for the interaction between two D-branes .
Thus, ${\widehat T}_p$ can be identified with the tension of
the D$p$-brane.
That this identification is correct can also be seen in a more
direct way. In fact, the boundary state (\ref{bounda}) is
the generator of the interaction amplitudes of
{\it any} closed string state with a D-brane. These amplitudes can be
simply computed by saturating $|B\rangle$ with the corresponding
normalized closed string states. In particular, the coupling between a
level 1 state of momentum $k$ and a D-brane is given by
~\footnote{We normalize each
component of the momentum eigenstates such that $\langle k|k'
\rangle = 2\pi\delta ( k-k')$ and take $(2\pi)^d\,\delta^{(d)}(0)=V$.}
\beq
A^{\mu \nu } \equiv \langle 0; k | a_{1}^\mu {\widetilde{a}}_{1}^\nu |
B\rangle = - \frac{{\widehat{T}}_{p}}{2}\, V_{p+1} \,{\cal S}^{\mu \nu}
\label{maless}~~,
\eeq
where $V_{p+1}$ is the world-volume of the D-brane.
The explicit expression of the emission amplitude for a graviton
and a dilaton
can then be obtained by saturating
$A^{\mu\nu}$ with the polarization tensors
\bea
\e_{\mu\nu}^{(h)} &=& \e_{\nu\mu}^{(h)}~~,
\hspace{1.5cm}
\e_{\mu\nu}^{(h)}\,\eta^{\mu\nu}=\e_{\mu\nu}^{(h)}k^\mu=0~~, \nl
\epsilon_{\mu \nu}^{(\phi)} &=& \frac{1}{\sqrt{d-2}}
\left[ \eta_{\mu \nu} - k_{\mu} \ell_{\nu} - k_{\nu}
\ell_{\mu}\right]~~, \hspace{1cm} k \cdot \ell =1,~~  \ell^2 =0
\label{dilpol}~~;
\ena
thus, using \eq{S} and the fact that the outgoing state has
non-vanishing momentum only along the transverse
directions, one obtains
\bea A_{grav} &=& A^{\mu\nu}\,\e_{\mu\nu}^{(h)}
= ~- {\widehat{T}}_{p} \,V_{p+1} \,\eta^{\alpha \beta}\,
\epsilon_{\alpha \beta}^{(h)}~~,\label{gradi}\\
A_{dil} &=& A^{\mu\nu}\,\e_{\mu\nu}^{(\phi)} =~
{\widehat{T}}_{p}\,V_{p+1} \,\frac{ d - 2p -4}{2 \sqrt{d-2}}~~.
\nonumber
\ena
Since $A^{\mu\nu}$ in \eq{maless} is symmetric, one can immediately
deduce that the coupling of the brane
with the antisymmetric Kalb-Ramond tensor
$B_{\mu\nu}$ is vanishing.
On the other hand, the interaction between the massless fields and a
D-brane is described by the Born-Infeld action~\footnote{$\kappa$ is
related to Newton's constant by $\kappa^2 = 8 \pi G_{N}$.}
\beq
S_{BI} = - \frac{T_{p}}{\kappa} \int d^{p+1} \xi ~{\rm e}^{- \phi}
\sqrt{- \det\left[ G_{\alpha \beta} + B_{\alpha \beta} +
2 \pi \alpha' F_{\alpha \beta} \right] }~~.
\label{borninfe}
\eeq
where $T_p$ is, by definition,
the D-brane tension and $F$ is the field strength
of an external gauge field. Notice that the action (\ref{borninfe}) is
written using the string metric $G$, and should be considered together
with the gravitational bulk action in the string frame
\beq
S_{bulk} = \frac{1}{2\kappa^2} \int d^{d} x \sqrt{ - G}
~{\rm e}^{- 2 \phi} \left[ R(G) + 4 ( \nabla \phi )^2
- \frac{1}{12} (dB)^2 \right]~~.
\label{bulk}
\eeq
This action, however, leads to a mixed propagator between
the dilaton and the graviton. To remove this mixed term, one has to
go to the Einstein frame and then rescale the dilaton field so that
its kinetic term is correctly normalized.
This can be done by writing
\beq
G_{\mu \nu} = {\rm e}^{4\phi/(d-2)} g_{\mu \nu}~~~,
~~~ \phi = \kappa\,\frac{\sqrt{d-2}}{2} \chi~~.
\label{ein}
\eeq
Using these definitions in Eq. (\ref{borninfe})
and putting for simplicity
$B=F=0$, we get
\beq
S_{BI} = - \frac{T_{p}}{\kappa} \int d^{p+1} \xi~{\rm e}^{- \kappa \chi
(d- 2p -4)/(2\sqrt{d-2})}
\sqrt{- \det\left[ g_{\alpha \beta} \right] }~~.
\label{borninfe2}
\eeq
If we consider the D$p$-brane in
the static gauge where $ x^{\alpha} \equiv
\xi^{\alpha}$, and expand the metric around the Minkowski background
$g_{\mu \nu} = \eta_{\mu \nu} + 2 \kappa  h_{\mu \nu}$,
it is not difficult to derive that the couplings of
the D-brane with the graviton and the dilaton are
\beq
A_{grav} = - T_p \,V_{p+1}\,
\eta^{\alpha \beta}\,\epsilon_{\alpha \beta}^{(h)}
~~~~,~~~~
A_{dil} = T_p\,V_{p+1}\,\frac{ d - 2p -4}{2 \sqrt{d-2}}~~.
\label{gradi2}
\eeq
Comparing Eqs. (\ref{gradi}) and (\ref{gradi2}), we deduce that
$T_p = {\hat{T}}_p$ in agreement with the result obtained by
 Polchinski~\cite{POLC}.
Notice that our calculation is shorter
than his because we know the precise
normalization of the boundary state and do not need to insert the
coupling just derived into a tree diagram
with the propagator of the massless
states in order to compare with the string calculation.

We now prove that the boundary state contains another
fundamental information about
the D-brane, namely the long-distance behavior
of the massless fields of the D-brane
solution of the string effective action. Before showing this, we recall
that an electric D$p$-brane is a solution of the field equations
derived from
\beq
S = \frac{1}{2 \kappa^2} \int d^d x \sqrt{-{\hat{g}}} \left[ {\hat{R}}
({\hat{g}})  -
\frac{{\hat{\gamma}}}{2} \left(\nabla {\hat{\phi}} \right)^2 -
\frac{1}{2 (n+1)!} {\rm e}^{- {\hat{a}}{\hat{\gamma}}{\hat{\phi}}}
\left(d{\hat{C}}_{(n)}\right)^{2}
\right]~~,
\label{action}
\eeq
where $n=p+1$ and ${\hat{\gamma}} = {2}/{(d-2)}$,
which corresponds to the ansatz
\beq
d {\hat{s}}^2 = \left[ H(x) \right]^{2 a}
\left(\eta_{\alpha\beta}dy^\a
dy^\b \right) + \left[ H(x) \right]^{2b} (\delta_{ij}d x^idx^j)~~,
\label{metri}
\eeq
for the metric ${\hat g}$, and
\beq
{\rm e}^{- {\hat{\phi}}(x)} = \left[ H(x) \right]^{\tau}
~~~~,~~~~{\hat{C}}_{01...p}(x) = \pm \sqrt{2 \sigma} [ H(x) ]^{-1}~~,
\label{dil}
\eeq
for the dilaton $\hat \phi$ and for the $(p+1)$-form potential
$\hat C$ respectively.
(The two signs in $\hat C$ correspond to
the brane and the anti-brane case.)
If the parameters are chosen as
\beq
a= - \frac{d - p -3}{2(d-2)}~~~,~~~b= \frac{p+1}{2(d-2)}~~~,~~~
\tau= \frac{{\hat{a}}}{2}~~~,~~~\sigma = \frac{1}{2}~~,
\label{expo}
\eeq
with ${\hat{a}}$ obeying the equation
\beq
(p+1)( d-p-3) + {\hat{a}}^2 = 2 (d-2)~~,
\label{equ}
\eeq
then the function $H(x)$ satisfies the flat space Laplace equation.
An extremal $p$-brane solution is constructed by introducing in the
right hand side a $\delta$-function source
term in the transverse directions.
If we restrict ourselves to the simplest case of just one
$p$-brane, we can write
\beq
H(x) = 1 + 2 \kappa T_p G(x) ~~,
\label{sol}
\eeq
where
\beq
G(x) = \left\{ \begin{array}{cc}
 \left[ (d-p -3) | x|^{(d-p-3)}
\Omega_{d-p-2}\right]^{-1} &  ~~~p < d-3 ~~,\\
  - \frac{1}{2 \pi} \log |x|  &   ~~~p= d-3 ~~,
\end{array} \right.
\label{solu}
\eeq
with $\Omega_{q}={ 2 \pi^{(q+1)/2}}/{ \Gamma \Big((q+1)/2\Big)}$
being the area of a unit $q$-dimensional sphere $S_q$.
Looking at the behavior for $|x|\to \infty$ of the fields
in Eqs. (\ref{metri}) and (\ref{dil}),
we see that they tend in general to non-vanishing background values.
At large distance, the fluctuations around these background values
(which we denote by ${\hat h}_{\mu\nu}$,
$\hat\varphi$ and ${\hat A}_{01...p}$ respectively)
correspond to the exchange
of the massless states: this is precisely
what is encoded in the boundary state.
To see this explicitly, it is first convenient to rescale the fields
according to (see {\it e.g.} \eq{RRact})
\beq
h_{\mu \nu} = \frac{1}{2 \kappa} {\hat{h}}_{\mu \nu} ~~~~,~~~~
\varphi = \frac{1}{ \kappa \sqrt{d-2}} {\hat{\varphi}} ~~~~,~~~~
A_{01...p} = \frac{\sqrt{2}}{\kappa}
{\hat A}_{01...p}~~, \label{newfi}
\eeq
and then make the Fourier transform
\beq
\int dt \, d^{p} y \,\, d^{d_{\bot}} x \,\,
{\rm e}^{ {\rm i} k_{\bot} \cdot x} G(x) = \frac{V_{p+1}}{{k}_{\bot}^{2}}
\label{genft}~~.
\eeq
The transformation (\ref{newfi}) is determined by requiring that, in
term of the new fields, the action (\ref{action}) becomes the one
usually obtained from the string calculations in the low energy limit.
Thus, to make easier the comparison with the string results, we
rewrite the classical solutions (\ref{metri}) and (\ref{dil}) as
\beq
h_{\mu \nu}(k) = 2 T_p \frac{V_{p+1}}{ k_{\bot}^2}\,
{\rm diag} \left(- a, a \dots a, b \dots b \right)~~,
\label{ft}
\eeq
and
\beq
\varphi(k)=  - \frac{{\hat{a}}}{\sqrt{d-2}}\,T_p\,
\frac{V_{p+1}}{k_{\bot}^{2}}
~~~~,~~~~
A_{01\dots p}(k) = \mp 2\sqrt{2}\,{T_p}
\frac{V_{p+1}}{k_{\bot}^2}
\label{dilft}~~.
\eeq
The classical solution has a mass per
unit $p$-volume, $M_p$ and an electric
charge with respect to the R-R field, $\mu_p$, given respectively by
\beq
M_p = \frac{T_p}{\kappa}~,~ \hspace{1cm} \mu_p = \pm \sqrt{2} T_p~~.
\label{char}
\eeq
The fact that the coefficient of the Coulomb-like behavior for the
gauge potential in \eq{dilft} is not exactly equal to $\mu_p$ is due
to the fact that the field $A$ of \eq{newfi} is not
"canonically" normalized.
The "canonically" normalized field is indeed $A^{can} =
A/2$, and in terms of it the coefficient of the
Coulomb-like term is exactly equal to $\mu_p$.

Now we will show that it is possible to derive the
long-distance behavior of the $p$-brane solution directly
from the boundary state, without making any reference to field
theory. In fact, a classical solution of the field equations can be
seen as a source for particles interacting with external quantum
states. The probability amplitude for this interaction to occur is
proportional to the Fourier transform of the classical field and is
related to the emission vertex of the various states.
In our case, such a vertex
is, as we have just shown, the boundary state.
However, it should be borne in mind that
the string description cannot be really continued off shell; thus we
can derive only
the most singular terms in $k_{\bot}$ which correspond
to the next-to-leading terms in the solutions (\ref{metri}) and
(\ref{dil}).
These can be obtained by projecting
the boundary state onto the level-one states with the
closed string propagator (\ref{pro}) inserted in between.
Simple algebra shows
that such a projection is
\beq
J^{\mu\nu}(k) \equiv
\langle 0; k | a_{1}^\mu  {\widetilde{a}}_{1}^\nu \,D_{a=1}| B\rangle =
- \frac{T_{p}}{2}\, \frac{V_{p+1}}{{{k}}_{\bot}^2}
\,{\cal S}^{\mu \nu} ~~,
\label{proj}
\eeq
where the momentum $k$ has non-vanishing components only in the
transverse directions.
If we decompose $J^{\mu\nu}(k)$ into its
irreducible components, we obtain a
traceless symmetric tensor $h_{\mu\nu}$,
a scalar $\varphi$ and an antisymmetric tensor $B_{\mu\nu}$,
which we can identify, respectively, with the
long-distance fluctuations of the
graviton, the dilaton and the Kalb-Ramond field
in the presence of a D-brane. In particular, using Eqs. (\ref{S})
(\ref{dilpol}), we have
\bea
h_{\mu\nu}(k) &\equiv& J_{\mu\nu}(k) -
\frac{J(k)\cdot\e^{(\phi)}}{\eta\cdot\e^{(\phi)}}\,\eta_{\mu\nu}\nl
&=& 2 T_p \frac{V_{p+1}}{{{k}}_{\bot}^2}\,
{\rm diag} \left(- a, a \dots a, b \dots b \right)~~,
\label{grav1} \\
\varphi(k) &\equiv& J(k) \cdot \e^{(\phi)} =
T_p  \frac{V_{p+1}}{{{k}}_{\bot}^{2}}
\frac{ d-2p-4}{2\sqrt{d-2}}~~,\label{dil1}\\
B_{\mu\nu}(k) &\equiv& \frac{1}{2}
\left(J_{\mu\nu}(k)-J_{\nu\mu}(k)\right)
= 0 ~~,\label{b1}
\ena
where $a$ and $b$ are given in \eq{expo}.

Comparing with Eqs. (\ref{ft}) and (\ref{dilft}),
we find perfect agreement
for the metric and the Kalb-Ramond field, which indeed was absent in the
$p$-brane solution; for the dilaton instead the agreement occurs
only if $d=10$. This strongly suggests to consider the superstring
theory. As a matter of fact, a comparison between the $p$-brane solution
of the action (\ref{action}) and a string calculation, does
make sense only in the
superstring case where the graviton, dilaton and Kalb-Ramond field
come from the
NS-NS sector and the antisymmetric
gauge potentials like $A_{\mu_1...\mu_{n}}$
from the R-R sector.
Nonetheless, the bosonic case we have considered in this section already
tells us what are the distinctive
features of the boundary state and how the
long-distance behavior of the massless fields is encoded in it.
In the next Sections we will consider in detail the superstring case
and get a complete mapping between the D$p$-brane solutions and the
boundary state calculations.

\vskip 1.5cm
\sect{The NS-NS sector of superstring}
\label{sez3}

In the superstring the boundary state is similar to that of
\eq{bounda} but contains an additional part depending on the
oscillators of the fermionic coordinates.
In the Neveu-Schwarz sector, the orbital part of the
GSO projected boundary state is~\cite{bounstate,BILLO}
\beq
| B\rangle_{NS} =
\frac{{T}_{p}}{2} ~
\delta^{(d_{\bot})} (q - y ) ~
\exp \!\left[ - \sum_{n=1}^{\infty} {a_{n}^{\mu}}^{\dagger}
{\cal S}_{\mu \nu} {{\widetilde{a}}_{n}^{\nu\dagger}}
\right] \,\sin \!\left[\sum_{r=1/2}^\infty {b_{r}^{\mu}}^{\dagger}
{\cal S}_{\mu \nu} {{\widetilde{b}}_{r}^{\nu\dagger}} \right]| 0;
k= 0\rangle~,
\label{bounda2}
\eeq
where $b_r^\mu $ $({\widetilde b}_r^\mu)$ are
the left (right) moving modes
of the NS fermionic string coordinates $ \psi^\mu $.
The normalization constant $T_p$ is given by
\eq{nor} with $d=10$.
However, to have a non-ambiguous
description, it is necessary
to specify the superghost charge or, equivalently, to
select a picture for the boundary state.
In our case, the simplest choice is to write
\beq
|{\widehat B}\rangle_{NS} = |B\rangle_{NS}~|\Omega_{NS}\rangle~~,
\label{bns}
\eeq
where
$|\Omega_{NS}\rangle $
is the superghost vacuum with charge $(-1,-1)$~\footnote{Here and
in the following, we explicitly write only the superghost vacuum and
understand the full ghost and superghost contributions to the
boundary state, since these do not play any
significant role for our present purposes.}. Of course, in
principle one is allowed to change this choice by applying
the usual picture changing
operator, but this modifies also the structure of the orbital
part (\ref{bounda2}).

As we have seen in the previous section,
in order to compare with the classical solution
we project the boundary state onto the massless states.
Of course, the latter have to be chosen in the picture dual to that
of the boundary state in order to soak up the superghost
anomaly on the disk. In NS-NS case these states are produced
by the vertex operator
\beq \label{VNS}
V^{NS\hbox{-}NS}(k;z,\bar z) =
\e_{\mu\nu} \,:{\cal V}_{-1}^\mu (k/2; z)
{\widetilde{\cal V}}_{-1}^\nu (k/2; \bar z):~~,
\eeq
with \footnote{Here and in the following we use
the notations of Ref.~\cite{FMS}:
$X^\mu(z)$ and $\psi^\mu(z)$ are the bosonic and fermionic string
coordinates, $b(z)$ and $c(z)$ are the ghost
fields, while $\phi(z)$, $\xi(z)$ and $\eta(z)$ are related to the
superghost fields.}
\beq \label{NS-1}
{\cal V}_{-1}^\mu (k; z) = {\rm e}^{-\phi(z)}\, c(z)\,\psi^\mu (z) \;
{\rm e}^{ik\cdot X(z)}~~.
\eeq
The antiholomorphic part, ${\widetilde{\cal V}}_{-1}^\nu$,
is also given by
\eq{NS-1} but with the left-moving fields replaced by the
corresponding right-moving ones ($X^\mu(z)\rightarrow
\widetilde X^\mu(\bar z)$, etc. etc.), and the polarization tensor
$\e_{\mu\nu}$ is as in \eq{dilpol}.

Thus, the projection one has to consider is
\beq
J^{\mu\nu}(k) \equiv \lim_{z,\bar z \to \infty}
\langle 0;0|
:{\cal V}_{-1}^\mu (k/2; z)
{\widetilde{\cal V}}_{-1}^\nu (k/2; \bar z):
D_{a=0} \,|{\widehat B}\rangle_{NS} ~~.
\label{projns}
\eeq
Computing the ghost and superghost contributions, one obtains
\beq
J^{\mu\nu}(k) = \langle 0; k| \,b_{1/2}^\mu {\tilde{b}}_{1/2}^\nu\,
D_{a= \frac{1}{2}}\,
| B\rangle_{NS} =
- \frac{T_p}{2} \frac{V_{p+1}}{{k}_{\bot}^{2}} \,{\cal S}_{\mu
\nu}~~,
\label{nsa}
\eeq
which is the same result of the bosonic string calculation, \eq{proj}.

The irreducible components of $J^{\mu\nu}(k)$ give the long-distance
behavior of the graviton, dilaton and Kalb-Ramond fields as
in Eqs. (\ref{grav1})--(\ref{b1}). Since now $d=10$,
this result is in complete agreement with the $p$-brane solution
in Eqs. (\ref{ft}) and (\ref{dilft}).

\vskip 1.5cm
\sect{The R-R sector of superstring}
\label{sez4}

The massless spectrum in the R-R sector of type II string theories
consists of a collection of
antisymmetric tensor fields $A_{\mu_1...\mu_{n}}$
playing the role of gauge
potentials. In the type IIA theory, these R-R fields are forms of
odd degree ($n=1,3,...$), while in the type IIB theory they are forms
of even degree ($n=0,2,...$)
\footnote{We understand that in ten dimensions a form of degree n
is equivalent to a form of degree $8-n$, since their field strengths
are dual to each other.}.
In the BRST invariant formalism, the emission of a R-R field from a
closed string is described by a vertex operator which usually is taken
to be
\beq \label{VRRa}
V^{R\hbox{-}R}(k;z,\bar z) = {\cal F}_{\alpha{\dot{\beta}}}
:{\cal V}_{-1/2}^{\alpha} (k/2;z)
\widetilde{\cal V}_{-1/2}^{\dot\beta} (k/2;\bar z):
\eeq
for type IIA, and
\beq \label{VRRb}
V^{R\hbox{-}R}(k;z,\bar z) = {\cal F}_{\dot\a \dot\b}
:{\cal V}_{-1/2}^{\dot\a} (k/2;z)
\widetilde{\cal V}_{-1/2}^{\dot\b} (k/2;\bar z):
\eeq
for type IIB.
Here we have adopted the conventions of Ref.~\cite{kos} and denoted by
$\a$ ($\dot\a$) the sixteen-dimensional indices of a chiral (antichiral)
Majorana-Weyl spinor of $SO(1,9)$.
The holomorphic component of the type IIA vertex operators is
\beq \label{RR-1/2}
{\cal V}_{-1/2}^\alpha (k;z) = {\rm e}^{-\phi(z)/2}\, c(z)\,
S^\alpha (z) \; {\rm e}^{ik\cdot X(z)}~~,
\eeq
where $S^\alpha (z)$ is the spin field ~\cite{FMS}. The antiholomorphic
component is given by analogous expressions in terms of the
right-moving fields.
The bispinor ${\cal F}_{\alpha{\dot\beta}}$,
which plays the same role of the polarizations
$\e_{\mu\nu}$ in the NS-NS vertices, is
\beq
{\cal F}_{\a\dot\b} =
\frac{{\rm i}}{16(n+1)!}\left(C\Gamma^{\mu_1\cdots \mu_{n+1}}
\right)_{\a\dot\b}\, F_{\mu_1...\mu_{n+1}} ~~,
\label{bispin}
\eeq
where $\Gamma^{\mu_1\cdots \mu_{n+1}}$ is the antisymmetrized
product of $(n+1)$ $\Gamma$-matrices and $C$ is the charge conjugation
matrix (see the Appendix for our conventions and notations).
Similar equations hold also for type IIB.
The BRST invariance of the vertices (\ref{VRRa}) and (\ref{VRRb})
requires $k^2=0$, $d*F=0$ and $dF=0$, so that $F$ must be
identified with a field strength ({\it i.e.} $F = dA$).

The vertices of Eqs. (\ref{VRRa}) and (\ref{VRRb})
have the same superghost charge in the left and right sectors,
like the NS-NS vertices (\ref{VNS}), and are the ones
that have been usually
considered in the literature
for the calculation of correlation functions.
However, this is not the only existing possibility.
For example, one could
choose to work with R-R vertices
that carry a total superghost charge $-2$
(which is the right amount to saturate the superghost anomaly on a 
disk \footnote{The necessity of using a 
R-R vertex in the asymmetric picture
with superghost number $(- 1/2, - 3/2)$ 
for obtaining a non vanishing emission
of a R-R state from the disk was pointed 
out in Ref.~\cite{BIANCHI}. The same
picture is also necessary for getting 
the central charges corresponding to the
R-R states in the supersymmetry algebra~\cite{NATHAN}.})
and that can be directly projected on a boundary state as we
did with the NS-NS vertices in the previous subsection.
R-R vertices of this kind can be easily constructed. For example, in
the type IIA theory, one can use the vertex operator
\bea \label{WRRa}
W^{R\hbox{-}R}(k;z,\bar z) &=&{\cal A}_{\dot{\alpha}\dot{\beta}}
:{\cal V}_{-3/2}^{\dot{\alpha}} (k/2;z)
\widetilde{\cal V}_{-1/2}^{\dot{\beta}} (k/2;\bar z):
\\
&+&{\rm i}\,4\sqrt{2} 
{\cal F}_{{\dot \a}\b}:{\cal V}_{-3/2}^{\dot{\alpha}}
(k/2;z) \widetilde{\cal V}_{-3/2}^{\beta} (k/2;\bar z)
{\bar\partial} {\widetilde c}(\bar z){\widetilde\xi}(\bar z) :~~,
\nonumber
\ena
with
\beq
{\cal A}_{\dot{\alpha}{\dot\beta}}= \frac{1}{4\sqrt{2}n!}
\left(C\Gamma^{\mu_1\cdots \mu_n}\right)_{{\dot\a}\dot\b}
~ A_{\mu_1...\mu_n} ~~~~~~{\hbox{($n$ odd)}}
\label{aslash}
\eeq
and
\beq \label{RR-3/2}
{\cal V}_{-3/2}^{\dot{\alpha}}(k;z) = {\rm e}^{-3\phi(z)/2}\, c(z)\,
S^{\dot{\alpha}} (z) \; {\rm e}^{ik\cdot X(z)}~~.
\eeq
In this case the BRST invariance implies $k^2=0$,
$d*A=0$ and $F=dA$, so that $A$ must be identified with
a gauge potential. Of course the vertices (\ref{WRRa}) and (\ref{VRRa})
are equivalent; this can be checked directly either by showing that
$W^{R\hbox{-}R}(k;z,\bar z)$ and
$V^{R\hbox{-}R}(k;z,\bar z)$ are related to each other
by a picture-changing operation in the left sector, or by showing that
they produce the same correlation functions.
In this respect it is worth pointing out that even if
$W^{R\hbox{-}R}(k;z,\bar z)$ contains the bare potential $A$, its
correlation functions
depend only on the field strength $F$; moreover, the second line
of \eq{WRRa} does not give any contribution inside expectation values,
but it is necessary for the BRST invariance of the whole vertex.
Notice that $W^{R\hbox{-}R}(k;z,\bar z)$
has the correct GSO projection for a vertex of the type IIA theory; in
fact even if the left and right spinor indices are in the same
spin representation, the different superghost charges
give the left and right
parts a different $F$-parity.
Of course a similar construction can be done also for the type IIB
with obvious changes in the spinor indices.

It is possible to check that the scattering amplitudes among the R-R
vertex operators (\ref{VRRa}),
(\ref{VRRb}), (\ref{WRRa}) and the NS-NS ones (\ref{VNS})
are reproduced, in the field theory limit, by the following
action
\beq
S' = \int d^{10}x\,\sqrt{-g}\left[-\frac{1}{8(n+1)!}~
{\rm e}^{\frac{5-n}{\sqrt{2}}\kappa\phi}
\left( dA_{(n)}\right)^2 \right]   ~~.
\label{RRact}
\eeq
By comparing this action with that in \eq{action}, used to
study the classical solution, one can easily explain the rescalings
in \eq{newfi}.

Let us now consider the interaction of the massless R-R fields with a
D-brane, and introduce the boundary state for the R-R
sector which can be used with the asymmetric vertex operators
(\ref{WRRa}). As explained in the Appendix,
this boundary state, after GSO projection, is
\bea
\label{BSR}
|{\widehat B}\rangle_{R} &=& \pm
\frac{T_p}{2}\delta^{d_\perp}(q-y)
\exp\!{\left[
  -\sum_{n=1}^\infty a_n^{\mu\dagger} {\cal S}_{\mu\nu}
  \widetilde{a}_n^{\nu\dagger}
  \right]} |0;\, k=0\rangle \\ \nonumber
&\times& \left\{ \cos\left[\sum_{n=1}^\infty
d_n^{\mu\dagger}{\cal S}_{\mu\nu}{\widetilde d}_n^{\nu\dagger}\right]
|\Omega_R\rangle^{(1)} +
\sin\left[\sum_{n=1}^\infty
d_n^{\mu\dagger}{\cal S}_{\mu\nu}{\widetilde d}_n^{\nu\dagger}\right]
|\Omega_R\rangle^{(2)} \right\} ~~,
\ena
with $d_n$ ($\widetilde{d}_n$) being the left (right)
moving modes of the fermionic coordinates and
\beq
|\Omega_R\rangle^{(1)} = \left\{\matrix{
M_{\a\b}\,|\a\rangle_{-1/2} |{\widetilde \b}\rangle_{-3/2}
~~~~~{\rm for~~~IIA~~}(p~\rm{even})\ \ , \cr \!\!
M_{\dot\a\b}\,|\dot\a\rangle_{-1/2} |{\widetilde \b}\rangle_{-3/2}
~~~~~{\rm for~~~IIB~~}(p~\rm{odd}) \ \ , \cr}\right.
\label{vacuum3}
\eeq
and
\beq
|\Omega_R\rangle^{(2)} =\left\{\matrix{
M_{\dot\a\dot\b}\,|\dot\a\rangle_{-1/2} |{\widetilde {\dot\b}}
\rangle_{-3/2}
~~~~~{\rm for~~~IIA~~}\ \ , \cr
M_{\a\dot\b}\,|\a\rangle_{-1/2} |{\widetilde {\dot\b}}\rangle_{-3/2}
~~~~~{\rm for~~~IIB~~}\ \ , \cr}\right.
\label{vacuum4}
\eeq
where we have introduced the notation
\beq
|\a\rangle_\ell \equiv \lim_{z \to 0} :S^\a(z)
{\rm e}^{\ell\phi(z)}:|0\rangle~~.
\label{astate}
\eeq
For a D$p$-brane the explicit expression for the matrix
$M$ to be used in Eqs. (\ref{vacuum3}) and (\ref{vacuum4}) is
\beq
M_{AB} \equiv \left( \begin{array}{cc} 
               M_{\alpha \beta} & M_{\alpha {\dot{\beta}}} \\
               M_{{\dot{\alpha}} \beta} & M_{{\dot{\alpha}} {\dot{\beta}}}
          \end{array} \right) =
 \left( C\Gamma^0\Gamma^1\cdots\Gamma^p\right)_{AB}~~.
\label{M}
\eeq

We first consider the type IIA theory and compute the interaction
between the R-R massless potential and a D-brane, following the
same procedure presented in \secn{sez2} for the bosonic case.
Saturating $|{\widehat B} \rangle_R$
with the corresponding state (\ref{WRRa}), namely
\beq
\lim_{z\to \infty} \langle 0 | W^{R\hbox{-}R}(k;z,\bar z) | {\widehat B}
\rangle_{R} = \mp\sqrt{2}T_p\,V_{p+1}\,A_{0...p}~~,
\eeq
we can reconstruct the Wess-Zumino term of the D-brane effective
action
\beq\label{actRR}
S_{WZ} = -\mu_p \int A~~,
\eeq
and derive again \eq{char}.

Now we turn to the R-R part of the solutions (\ref{dil}) and insert a
closed string propagator in the boundary state projection
\bea
J^{{\dot\a}{\dot\b}}(k) &\equiv& \lim_{z,\bar z \to \infty}
\langle 0|:
{\cal V}_{-3/2}^{\dot\a} (k/2; z)
{\widetilde{\cal V}}_{-1/2}^{\dot\b} (k/2; \bar z):
D_{a=0}| {\widehat B}\rangle_R
\nonumber \\
&=&\mp \frac{T_p}{2}\,\frac{V_{p+1}}{k_\perp^2}\,
\left(C^{-1}MC^{-1}\right)^{{\dot\a}{\dot\b}} ~~,
\label{projr}
\ena
Then, we decompose $J^{{\dot\a}{\dot\b}}$ into irreducible
components to obtain the gauge potentials.
Since \eq{aslash} instructs us to identify the trace of the
bispinor with the potential $A$ times a factor of $2\sqrt{2}$, we
have
\beq
A_{\mu_1...\mu_n}(k)=\frac{1}{2\sqrt{2}}
{\rm Tr}\left(J(k)C\Gamma_{\mu_1}\cdots\Gamma_{\mu_n}\right)~~.
\eeq
Using \eq{projr},
we see that the potentials above are zero unless $n=p+1$, in which case
we get
\beq
A_{\mu_1...\mu_{p+1}}(k) =\mp
2\sqrt{2}\,{T_p}\,\frac{V_{p+1}}{k_\perp^2}
\,\varepsilon^{(v)}_{\mu_1...\mu_{p+1}}~~,
\label{Ap}
\eeq
where $\varepsilon^{(v)}$ is the completely antisymmetric tensor of
the world-volume.
This result is in perfect agreement with the classical solution
in \eq{dilft}.
The vanishing of all other potentials is consistent
with the fact that a D$p$-brane is charged only under the $(p+1)$ gauge
field of the R-R sector.

Similar calculations also hold for type
IIB theory and, also in this case,
we correctly reproduce the long-distance behavior of the
classical solution in agreement with the results of Ref. \cite{Myers}.

\vskip 1.5cm
\sect{Delocalized D-brane bound states}
\label{sez5}

We now show that, using the same procedure described in the previous
sections, it is possible to derive
the long-distance behavior of a broader
class of $p$-branes solutions; in particular we examine those background
field configurations that present two non trivial R-R potentials whose
dimension differs by two.
These solutions are described in detail in Ref.~\cite{Myers2}, where
they are explicitly constructed from those that present only one
form field potential using the T-duality symmetry between type IIA and
IIB theories.
It is useful to outline here this approach since it can be directly
applied at a string level, obtaining a generalization of the boundary
state presented in the previous sections.
For simplicity, we again use
the boundary state of the bosonic string given in
Eq. (\ref{bounda0}). The same procedure can then be generalized in a
straightforward way to the superstring.

The first step of the procedure presented in Ref.~\cite{Myers2}
is to delocalize the $p$-brane solution removing its dependence also
on the $(p+1)^{\rm th}$ spatial direction; from the world-sheet point of
view this operation simply consists in fixing to zero
the momentum $k_{p+1}$ along a Dirichlet direction
so that, in the final
result, the square of transverse momentum
does not involve $k_{p+1}$, that is
${k}_\bot^2 = \sum\limits_{i=p+2}^d k_i k^i$.
The boundary state corresponding
to this delocalized solution is again given
by the same expression as in Eq.(\ref{bounda0}) with, however,
two important modifications. One is the
absence of the $\delta$-function along the delocalized
direction $(p+1)$ and
the other is a different normalization:
in fact, according to Eq. (\ref{norm1}), one $\delta$-function
less implies also a factor of $( 2 \pi \sqrt{\alpha'})$ less,
and this corresponds to substitute $T_p$
with $T_{p+1}$ in the normalization. In conclusion the boundary state
corresponding to a D$p$-brane delocalized in
the $(p+1)^{\rm th}$ direction
is given by
\beq
|B\rangle =  \frac{T_{p+1}}{2} ~
\delta^{(d_{\bot}')} (q - y) ~
\exp \left[ - \sum_{n=1}^{\infty} a_n^{\mu\dagger}
{\cal S}_{\mu \nu} {\widetilde a}_n^{\nu\dagger}
\right]  | 0; k= 0\rangle
\label{debounda}~~,
\eeq
where $d_{\bot}'$ means that the
$\delta$-function in the $(p+1)^{\rm th}$
direction is absent.

Now, in the plane defined by
the delocalized Dirichlet direction and
by the $p^{\rm th}$ Neumann direction,
we perform first a rotation with angle $\phi$ and then a T-duality
transformation along
the $(p+1)^{\rm th}$ transverse coordinate axis. These operations do not
affect the structure of the boundary
state (\ref{debounda}) but simply that
of matrix ${\cal S}_{\mu\nu}$. In fact, under the rotation
${\cal S}_{\mu\nu}$ behaves
as a tensor, while the T-duality changes the sign of the component of
the $(p+1)^{\rm th}$ column, yielding
\beq
{\cal S}_{\mu\nu}= \left\{ \begin{array}{cl}
\eta_{\mu\nu} &\;\;\;\hbox{if}\;\;\mu=0,\ldots,p-1~~,
\\ \label{Snew}
-\delta_{\mu\nu} &\;\;\;\hbox{if}\;\;\mu=p+2,\ldots,d-1~~,\\
\left(\matrix{ \cos 2\phi & -\sin 2\phi\cr
\sin 2\phi & \cos 2\phi\cr }\right)
&  \;\;\; \hbox{in the plane}\;\; p\,,p+1~~.
\end{array}
\right.
\eeq

The rotation can be performed with the same procedure used in
Ref.~\cite{BILLO} for boosting the boundary state. There is, however, a
very important difference with respect to the boost case.
Unlike in the case
discussed in
Ref.~\cite{BILLO} in this case
there is no $\delta$-function in the transverse
direction in which we perform the rotation
and therefore the rotation acts
trivially on the zero modes. As a result the rotation in a delocalized
D$p$-brane does not generate any
``Born-Infeld'' factor as instead happens in
the localized case.

With this new matrix appearing in the exponent the boundary state
satisfies new overlap equations in the plane where we
performed the rotation
\bea \label{bcond}
&\left( \partial_{\tau} X^p \;
+{\rm i}\, \tan\phi\;\partial_{\sigma} X^{p+1}
\right)|_{\tau=0} |B \rangle  & =  0~~,
\\ \nonumber
& \left( \partial_{\tau} X^{p+1}
-{\rm i}\, \tan\phi\;\partial_{\sigma} X^{p}
\right)|_{\tau=0} |B \rangle & = 0~~.
\ena
Those are the same overlap equations satisfied by a boundary state
in presence of a costant background
field ${\cal F}=B+2\pi\a' F$, where $B$
is the Kalb-Ramond potential
and $F$ is the external gauge field strength:
\beq
\left( \partial_{\tau} X^{\mu}
-{\rm i}\, {\cal{F}}^{\mu \nu} \partial_{\sigma} X_{\nu}
\right)_{\tau=0} |B \rangle =0 ~~,
\label{extfield}
\eeq
where $\mu, \nu = p$ or $p+1$.
Comparing Eqs. (\ref{bcond}) and (\ref{extfield})
we get that
\beq
{\cal F}_{(p+1)\,p} = -{\cal F}_{p\,(p+1)} =\tan\phi ~~.
\label{rel}
\eeq
The normalization of the boundary
state of a delocalized $p$-brane differs,
however, from the one with an external field because of the lack, in the
case of a delocalized $p$-brane,
of the Born-Infeld factor present instead
in the boundary state with an external field.

Now we have lifted the whole construction of Ref.~\cite{Myers2} at a
string level. Thus, using the same procedure of
previous sections, we can show that the new boundary state is
the conformal description of those delocalized  $p$-branes that present
two non-vanishing R-R potentials.

The above construction of a delocalized
boundary state for the bosonic string
can be easily generalized to the NS-NS sector
of the superstring as we have
done in sect. \ref{sez2}. In particular,
the boundary state for the NS-NS sector
of a delocalized $p$-brane has exactly the same form as the one in
Eq. (\ref{bounda2}) with the matrix ${\cal{S}}$
given in Eq. (\ref{Snew}), with
no $\delta$-function in the $(p+1)^{\rm th}$
transverse direction, and with $T_p$ replaced by $T_{p+1}$. Then, using
\eq{nsa}, from this boundary state
one can obtain the correct large distance
behavior of the NS-NS fields of this type of solutions
\bea
h_{\mu\nu}(k) &=&
2T_{p+1} \,
\frac{V_{p+2}}{{k}_\bot^2}\,{\rm diag}(-a,a,...,a,c,c,b,...,b)~~,
\label{h2} \\
\varphi(k) &=& -{T_{p+1}}\,\frac{V_{p+2}}{{k}_\bot^2}\,
\frac{p-3+\cos2\phi}{2\sqrt{2}}~~,\label{dil2}\\
B_{p\,(p+1)}(k) &=& -B_{(p+1)\,p}(k) =
\frac{T_{p+1}}{2} \,\frac{V_{p+2}}{{k}_\bot^2} \sin 2\phi
 ~~,\label{b2}
\ena
where $a=(p+\cos2\phi -7)/16$, $c=(p-3(1+\cos2\phi))/16$, and
$b= (p+1 + \cos 2\phi)/16$. Remember that $k_{\bot}^{2}$ does not
contain its transverse component along the $(p+1)^{\rm th}$ direction.

The boundary state of a delocalized $p$-brane for the R-R sector has the
same form as the one in Eq. (\ref{BSR}) suitably modified as we have done
in the case of the NS-NS sector.
In addition, in the R-R sector the rotation
and the T-duality transformation act
also on the matrix $M$ present in the
vacua of Eqs. (\ref{vacuum3}) and (\ref{vacuum4}).
Since a rotation in the $(p,p+1)$ plane
acts on the $\Gamma$-matrices as follows
\beq
\left( \begin{array}{c} \Gamma^{p} \\
                        \Gamma^{p+1}
     \end{array} \right)  \rightarrow
                   \left( \begin{array}{cc} \cos \phi & \sin \phi \\
                                             - \sin \phi & \cos \phi
 \end{array} \right)\left( \begin{array}{c} \Gamma^{p} \\
                        \Gamma^{p+1}
     \end{array} \right)
\label{act}
\eeq
and a T-duality transformation in the
$(p+1)^{\rm th}$ direction amounts to
multiply $M$ with $\Gamma^{p+1}$ from the right (see also \eq{defM'}),
then the matrix $M$ becomes~\footnote{Alternatively one
can deduce the new form for the matrix $M$ by
requiring it to satisfy \eq{eqM} but with the new
$S_{\mu\nu}$ given in \eq{Snew}.}
\beq
M = C\Gamma^0\cdots\Gamma^{p-1}(\sin\phi + \cos\phi\; \Gamma^p
\Gamma^{p+1})~~.
\label{Mbis}
\eeq
Since the the matrix $M$ now contains two terms
with a different number of
$\Gamma$-matrices, it is easy to realize that
\eq{projr} yields {\it two}
non-vanishing antisymmetric gauge potentials in the R-R sector, namely
\bea
A_{0...p-1}(k) & = &
\mp 2\sqrt{2}\,T_{p+1}\,\frac{V_{p+2}}{k_\perp^2}\,\sin\phi
~~,
\label{Ap1}\\ \label{Ap2}
A_{0...p+1}(k) & = &
\mp 2\sqrt{2}\,T_{p+1}\,\frac{V_{p+2}}{k_\perp^2}\, \cos\phi
~~.
\ena
The simultaneous appearance of a $(p-1)$-form
and a $(p+1)$-form in the R-R sector
indicates the presence of a D$(p-1)$-brane
and a D$(p+1)$-brane forming a bound
state, whose world-sheet realization is
given by the rotated and T-dualized
boundary state. As we have shown, this boundary state correctly
reproduces the long-distance behavior of the
delocalized D-brane classical solutions found in Ref.~\cite{Myers2}.

\vskip 1cm
\noindent
{\large{\bf {Acknowledgements}}}
\vskip 0.5cm
\noindent
We thank M. Bill{\'{o}} and D. Cangemi for very useful discussions.

\vskip 1.5cm
%\newpage
\appendix{\Large {\bf {Appendix A}}}
\label{appa}
\vskip 0.5cm
\renewcommand{\theequation}{A.\arabic{equation}}
\setcounter{equation}{0}
\noindent
Let $\gamma^i$ be the eight $ 16 \times 16 $ $\gamma$-matrices of
$SO(8)$. Starting from these matrices we can construct a chiral
representation for the $ 32 \times 32 $ $\Gamma$-matrices of
$SO(1,9)$, {\it i.e.}
\bea
\Gamma^{i} &=&
\pmatrix{
0 & \gamma^i \cr
\gamma^i & 0 \cr
} = \sigma^1 \otimes
\gamma^i  ~~,
\nonumber \\
\Gamma^{9} &=&
\pmatrix{
0 & \gamma^1\cdots\gamma^8 \cr
\gamma^1\cdots\gamma^8 & 0 \cr
} = \sigma^1 \otimes \left(\gamma^1\cdots\gamma^8 \right)~~,
\label{gamma} \\
\Gamma^{0} &=&
\pmatrix{
0 & \one \cr
-\one & 0 \cr
} = {\rm i} \,\sigma^2 \otimes \one   ~~,
\nonumber
\ena
where $\sigma^a$'s are the standard Pauli matrices.
One can easily verify that these matrices satisfy
$\{\Gamma^\mu\,,\, \Gamma^\nu\}=2\eta^{\mu\nu}$.
Other useful matrices are
\bea
\Gamma_{11} &=& \Gamma^0\ldots \Gamma^9=\pmatrix{
\one & 0 \cr
0 & -\one \cr
} = \sigma^3 \otimes \one  ~~,
\label{CC} \\
C &=& \pmatrix{
0 & -{\rm i}\one \cr
{\rm i}\one & 0 \cr
} = \sigma^2 \otimes \one ~~,
\nonumber
\ena
where $C$ is the charge conjugation matrix such that
\beq
\left(\Gamma^\mu\right)^T = - C\,\Gamma^\mu\,C^{-1}  ~~.
\label{transp}
\eeq

Let $A,B,...$ be 32-dimensional
indices for spinors in ten dimensions, and
$|A\rangle |{\widetilde B}\rangle$ denote the vacuum of
the Ramond fields $\psi^\mu(z)$ and
${\widetilde \psi}^\mu({\bar z})$ with spinor indices $A$ and
$B$ in the left and right sectors respectively, that is
\beq
|A\rangle |{\widetilde B}\rangle =
\lim_{z,{\bar z}\to 0} S^A(z)\,{\widetilde S}^B({\bar z})
|0\rangle
\label{vacapp}
\eeq
where $S^A$ (${\widetilde S}^B$) are the left (right)
spin fields \cite{FMS}, and $|0\rangle$ the Fock vacuum of the
Ramond fields.
The action of the Ramond oscillators $d_n^\mu$ and ${\widetilde
d}_n^\mu$ on the state $|A\rangle |{\widetilde B}\rangle$ is given by
\beq
d_n^\mu \,|A\rangle |{\widetilde B}\rangle
={\widetilde d}_n^\mu\,
|A\rangle |{\widetilde B}\rangle
= 0
\label{dn}
\eeq
if $n>0$, and
\bea
d_0^\mu\, |A\rangle |{\widetilde B}\rangle
&=& \frac{1}{\sqrt{2}} \left(\Gamma^\mu\right)^A_{~C}
\,\left(\!\one\, \right)^B_{~D}|C\rangle\, |{\widetilde D}\rangle
\nl
{\widetilde d}_0^\mu \,|A\rangle |{\widetilde B}\rangle
&=& \frac{1}{\sqrt{2}} \left(\Gamma_{11}\right)^A_{~C}
\,\left(\Gamma^\mu\right)^B_{~D}\,
|C\rangle |{\widetilde D}\rangle
\label{d0}
\ena
It is easy to check that this action correctly
reproduces the anticommutation properties of the $d$-oscillators,
and in particular that
$\{d_0^\mu\,,\,d_0^\nu\}=\{{\widetilde d}_0^\mu
\,,\,{\widetilde d}_0^\nu\}=\eta^{\mu\nu}$, and
$\{d_0^\mu\,,\,{\widetilde d}_0^\nu\}=0.$

We now use these definitions to derive the fermionic structure of the
boundary state $|B\rangle$ in
the R-R sector of a D$p$-brane.
According to the general theory \cite{NBOUND},
$|B\rangle$ has to satisfy
the following overlap equations
\beq
\left( d_n^\mu -{\rm i} \,{\cal S}_\nu^\mu {\widetilde d}_{-n}^\nu
\right)\,|B\rangle= 0
\label{overlap}
\eeq
where ${\cal S}_\nu^\mu$ is the matrix defined in \eq{S}.
For $n\not = 0$, \eq{overlap} is easily
solved by a Bogoliubov transformation;
indeed, the non-zero mode part of $|B\rangle$ is
\beq
|B\rangle' = \exp\! \left[{\rm i}\sum_{n=1}^\infty
d_n^{\mu\dagger}{\cal S}_{\mu\nu}{\widetilde d}_n^{\nu\dagger}\right]
\,|0\rangle
\label{bnonzero}
\eeq
where, as usual, $d_n^\dagger\equiv d_{-n}$.
The zero-mode part $|B\rangle^{0}$
requires more care due to the non trivial action
of the oscillators $d_0^\mu$ and ${\widetilde d}_0^\mu$.
If we write
\beq
|B\rangle^{0} = {\cal M}_{AB}\,|A\rangle |{\widetilde B}\rangle
\label{appB0}
\eeq
then, \eq{overlap} for $n=0$ implies that the $32\times32$ matrix
${\cal M}$ has to satisfy the following equation
\beq
\left(\Gamma^\mu\right)^T\,{\cal M} - {\rm i}\,
{\cal S}_\nu^\mu\,\Gamma_{11}\,{\cal M}\,\Gamma^\nu = 0
\label{eqM}
\eeq
Using our previous definitions, one finds that a solution is
\beq
{\cal M} = C\,\Gamma^0\cdots\Gamma^p\,\frac{1+{\rm i}
\,\Gamma_{11}}{1+{\rm i}}
\label{defM}
\eeq
If one performs a T-duality transformation in one of the transverse
directions of the D-brane, say for example in the $i^{\rm th}$ direction,
the boundary state retains its
structure but one must change the signs in the $i^{\rm th}$ column
of ${\cal S}_{\mu\nu}$
and replace ${\cal M}$ with the matrix
\beq
{\cal M'} = {\rm i}\,{\cal M}\,\Gamma^i\,\Gamma_{11} =
C\,\Gamma^0\cdots\Gamma^{p}\Gamma^i\,
\frac{1+{\rm i}\,\Gamma_{11}}{1+{\rm i}} 
\label{defM'}
\eeq

In our representation of $\Gamma$-matrices, it is natural
to decompose the spinors in chiral and antichiral components ($A=(\alpha,
{\dot{\alpha}}$)) with sixteen-dimensional indices $\a$ and $\dot\a$ 
respectively.
From Eqs. (\ref{gamma}) and (\ref{CC}), it follows
that for $p$ even ({\it{i.e.}} in the
type IIA theory), the matrix ${\cal M}$
has non-vanishing entries only in the diagonal blocks, that is in
the chiral-chiral sector and the antichiral-antichiral one, whereas for
$p$ odd ({\it{i.e.}} in the type IIB theory)
${\cal M}$ is non-trivial only in the off-diagonal
blocks, that is in the antichiral-chiral
sector and in the chiral-antichiral one.
Thus, in the sixteen-dimensional notation, \eq{appB0} becomes
\beq
|B\rangle^0 = |\Omega_R\rangle^{(1)} -{\rm i}\,
|\Omega_R\rangle^{(2)}
\label{vacuumap}
\eeq
where
\beq
|\Omega_R\rangle^{(1)} =\left\{\matrix{
M_{\a\b}\,|\a\rangle_{-1/2} |{\widetilde \b}\rangle_{-3/2}
~~~~~{\rm for~~~IIA~~}\ \ , \cr
M_{\dot\a\b}\,|\dot\a\rangle_{-1/2} |{\widetilde \b}\rangle_{-3/2}
~~~~~{\rm for~~~IIB~~}\ \ , \cr}\right.
\label{ome1}
\eeq
and
\beq
|\Omega_R\rangle^{(2)} =\left\{\matrix{
M_{\dot\a\dot\b}\,|\dot\a\rangle_{-1/2} |{\widetilde {\dot\b}}
\rangle_{-3/2}
~~~~~{\rm for~~~IIA~~}\ \ , \cr
M_{\a\dot\b}\,|\a\rangle_{-1/2} |{\widetilde {\dot\b}}\rangle_{-3/2}
~~~~~{\rm for~~~IIB~~}\ \ , \cr}\right.
\label{ome2}
\eeq
In these equations we have defined
\beq
M_{AB} \equiv \left( \begin{array}{cc} 
M_{\alpha \beta} & M_{\alpha {\dot{\beta}}} \\
M_{{\dot{\alpha}} \beta} & M_{{\dot{\alpha}} {\dot{\beta}}}
\end{array} \right) = \left(C\,\Gamma^0\cdots \Gamma^{p}
\right)_{AB}~~,
\eeq
and introduced the appropriate superghost charge by treating
asymmetrically the
left and right sectors as we explained in \secn{sez4}.
Then, the boundary state becomes
\beq
|B\rangle = \exp\!\left[{\rm i}\sum_{n=1}^\infty
d_n^{\mu\dagger}{\cal S}_{\mu\nu}{\widetilde d}_n^{\nu\dagger}\right]
\,|B\rangle^0
\label{b00}
\eeq
Finally, by applying on it
the GSO projection, we obtain
\bea
|{\widehat B}\rangle_R &=&
\frac{1-(-1)^p(-1)^F}{2}\,\frac{1+
(-1)^{\widetilde F}}{2}|B\rangle
\label{BGSO} \\
&=& \cos\left[\sum_{n=1}^\infty
d_n^{\mu\dagger}{\cal S}_{\mu\nu}{\widetilde d}_n^{\nu\dagger}\right]
|\Omega_R\rangle^{(1)} +
\sin\left[\sum_{n=1}^\infty
d_n^{\mu\dagger}{\cal S}_{\mu\nu}{\widetilde d}_n^{\nu\dagger}\right]
|\Omega_R\rangle^{(2)}
\nonumber
\ena
where $F$ and $\widetilde F$ are the left and right
F-parity operators defined in Ref.~\cite{kos}
which measure the fermion number, the chirality and the
superghost charge. Note that the GSO projection
in \eq{BGSO} is of type IIA for $p$ even
and of type IIB for $p$ odd in accordance with the R-R charge
that is carried by the D$p$-brane.

\end{document}